# A description of the transverse momentum distributions of charged particles produced in heavy ion collisions at RHIC and LHC energies

Jia-Qi Hui, Zhi-Jin Jiang[*], and Dong-Fang Xu

*College of Science, University of Shanghai for Science and Technology, Shanghai 200093, China*

[*]jzj265@163.com

By assuming the existing of memory effects and long-range interactions in the hot and dense matter produced in high energy heavy ion collisions, the nonextensive statistics together with the relativistic hydrodynamics including phase transition is used to discuss the transverse momentum distributions of charged particles produced in heavy ion collisions. It is shown that the combined contributions from nonextensive statistics and hydrodynamics can give a good description to the experimental data in Au+Au collisions at $\sqrt{s_{NN}} = 200$ GeV and in Pb+Pb collisions at $\sqrt{s_{NN}} = 2.76$ TeV for $\pi^\pm$, $K^\pm$ in the whole measured transverse momentum region, and for $p(\bar{p})$ in the region of $p_T \leq 2.0$ GeV/c. This is different from our previous work, where, by using the conventional statistics plus hydrodynamics, the describable region is only limited in $p_T \leq 1.1$ GeV/c.

*Keywords*：Nonextensive statistics; relativistic hydrodynamics; phase transition; transverse momentum distribution

## 1. Introduction

The primary goals of experimental programs performed in high energy heavy ion collisions are to find the deconfined nuclear matter, namely the quark-gluon plasma (QGP), which is believed to have filled in the early universe several microseconds after the big bang. Therefore, studying the properties of QGP is important for both particle physics and cosmology. In the past decade, a number of bulk observables about charged particles, such as the Fourier coefficients $v_n$ of azimuth-angle distributions [1, 2], transverse momentum spectra [3-8] and pseudorapidity distributions [9-12], have been extensively studied in nuclear collisions at both RHIC and LHC energies. These investigations have shown a fact that the matter created in these collisions is in the state of strongly coupled quark-gluon plasma (sQGP) exhibiting a clear collective behavior nearly like a perfect fluid with very low viscosity [13-31]. Therefore, the movement of sQGP can be described in the scope of relativistic hydrodynamics which connects the static aspects of sQGP and the dynamical aspects of heavy ion collisions [32].

In our previous work [33], by considering the effects of thermalization, we once used a hydrodynamic model incorporating phase transition in analyzing the transverse momentum distributions of identified charged particles produced in heavy ion collisions. In that model, the quanta of hot and dense matter are supposed to observe the standard statistical distributions and the experimental measurements in Au+Au collisions at $\sqrt{s_{NN}} = 200$ and 130 GeV can be well matched up in the region of $p_T \leq 1.1$ GeV/c. Known from the investigations in Ref. [34, 35], the memory effects and long-range force might appear in the hot and dense matter. This guarantees, at least to a certain extent, the reasonableness of nonextensive statistical approach in describing the thermal motions of quanta of hot and dense matter. Hence, in this paper, on the basis of hydrodynamics taking phase transition into considerations, we will use nonextensive statistics instead of conventional statistics to simulate the transverse collective flow of the matter created in collisions.

The nonextensive statistics is also known as Tsallis nonextensive thermostatistics, which was first proposed by C. Tsallis in 1988 in his pioneering work [36]. This statistical theory overcomes the shortcomings of the conventional statistics in many physical problems with long-range interactions, long-range microscopic memory, or fractal space-time constrains. It has a wide range of applications in astrophysical self-gravitating systems [37], cosmology [38], the solar neutrino problem [39], many-body theory, dynamical linear response theory and variational methods [40].

In the following section 2, a brief description is given to the adopted hydrodynamics, presenting its analytical solutions. The solutions are then used in section 3 to formulate the transverse momentum distributions of charged particles produced in heavy ion collisions in the light of nonextensive statistics and Cooper-Frye prescription. The last section 4 is about conclusions.

## 2. A brief introduction to the hydrodynamic model

The key points of the hydrodynamic model [18] used in the present paper are as follows.

The expansions of fluid follow the continuity equation

$$\frac{\partial T^{\mu\nu}}{\partial x^\nu} = 0, \quad \mu, \nu = 0, 1, \tag{1}$$

where $x^\nu = (x^0, x^1) = (t, z)$, and

$$T^{\mu\nu} = (\varepsilon + p)u^\mu u^\nu - p g^{\mu\nu}, \tag{2}$$

is the energy-momentum tensor of perfect fluid, where $g^{\mu\nu} = \text{diag}(1, -1)$ is the metric tensor, $u^\mu = (u^0, u^1) = (\cosh y_F, \sinh y_F)$ is the four-velocity of fluid with rapidity $y_F$. $\varepsilon$ and $p$ are the energy density and pressure of fluid, which are related by the sound speed of fluid $c_s$ via the equation of state

$$\frac{dp}{d\varepsilon} = \frac{sdT}{Tds} = c_s^2, \tag{3}$$

where, $T$ is the temperature, and $s$ is the entropy density of fluid.

Project Eq. (1) to the direction of $u_\mu$ giving

$$\frac{\partial(su^\nu)}{\partial x^\nu} = 0, \tag{4}$$

which is the continuity equation for entropy conservation. Project Eq. (1) to the direction perpendicular to $u_\mu$ leading to equation

$$\frac{\partial(T \sinh y_F)}{\partial t} + \frac{\partial(T \cosh y_F)}{\partial z} = 0, \tag{5}$$

which means the existence of scalar function $\phi$ satisfying

$$\frac{\partial \phi}{\partial t} = T \cosh y_F, \quad \frac{\partial \phi}{\partial z} = -T \sinh y_F. \tag{6}$$

From $\phi$ and Legendre transformation, Khalatnikov potential $\chi$ is introduced by

$$\chi = \phi - tT \cosh y_F + zT \sinh y_F, \tag{7}$$

which makes the coordinates of $(t, z)$ transform to

$$t = \frac{e^\omega}{T_0}\left(\frac{\partial \chi}{\partial \omega} \cosh y_F + \frac{\partial \chi}{\partial y_F} \sinh y_F\right), \tag{8}$$

$$z = \frac{e^\omega}{T_0}\left(\frac{\partial \chi}{\partial \omega} \sinh y_F + \frac{\partial \chi}{\partial y_F} \cosh y_F\right), \tag{9}$$

where $T_0$ is the initial temperature of fluid, and $\omega = \ln(T_0/T)$. In terms of $\chi$, Eq. (4) can be rewritten as the so-called telegraphy equation

$$\frac{\partial^2 \chi}{\partial \omega^2} - 2\beta \frac{\partial \chi}{\partial \omega} - \frac{1}{c_s^2}\frac{\partial^2 \chi}{\partial y_F^2} = 0, \quad \beta = \frac{1 - c_s^2}{2c_s^2}. \tag{10}$$

With the expansions of created matter, its temperature becomes lower and lower. When the temperature reduces from initial temperature $T_0$ to critical temperature $T_c$, the matter transforms from the sQGP state to the hadronic state. The produced hadrons are initially in the violent and frequent collisions, which are mainly inelastic. Hence, the abundance of identified hadrons is in changing. Furthermore, the mean free paths of these primary hadrons are very short. The evolution of them still satisfies Eq. (10) with only difference being the values of $c_s$. In sQGP, $c_s = c_0 = 1/\sqrt{3}$. In the hadronic state, $0 < c_s = c_h \leq c_0$. At the point of phase transition, $c_s$ is discontinuous.

The solutions of Eq. (10) for the sQGP and hadronic state are respectively [18],

$$\chi_0(\omega, y_F) = \frac{Q_0 c_0}{2} e^{\beta_0 \omega} I_0\left(\beta_0 \sqrt{\omega^2 - c_0^2 y_F^2}\right), \tag{11}$$

$$\chi_h(\omega, y_F) = \frac{Q_0 c_0}{2} S(\omega) I_0[\lambda(\omega, y_F)], \tag{12}$$

where $Q_0$ is a constant determined by fitting the theoretical results with experimental data, $I_0$ is the 0th order modified Bessel function, and

$$\beta_0 = (1 - c_0^2)/2c_0^2 = 1, \ S(\omega) = e^{\beta_h(\omega - \omega_c) + \beta_0 \omega_c}, \ \lambda(\omega, y_F) = \beta_h c_h \sqrt{y_h^2(\omega) - y_F^2},$$

$$\beta_h = (1 - c_h^2)/2c_h^2, \ \omega_c = \ln(T_0/T_c), \ y_h(\omega) = [(\omega - \omega_c)/c_h] + (\omega_c/c_0).$$

### 3. The transverse momentum distributions of charged particles produced in heavy ion collisions

**(1) The energy of quantum of produced matter**

In the nonextensive statistics, there are two basic postulations [36, 39]

(a) The entropy of a statistical system possesses the form of

$$s_q = \frac{1}{q-1} \sum_{i=1}^{\Omega} p_i(1 - p_i^{q-1}), \tag{13}$$

where $p_i$ is the probability of a given microstate among $\Omega$ different ones and $q$ is a real parameter.

(b) The mean value of an observable $O$ is given by

$$\bar{O}_q = \sum_{i=1}^{\Omega} p_i^q O_i, \tag{14}$$

where $O_i$ is the value of an observable $O$ in the microstate $i$.

From above two postulations, the average occupational number of quantum in the state with temperature $T$ can be written in a simple analytical form [41]

$$\bar{n}_q = \frac{1}{[1+(q-1)(E-\mu_B)/T]^{1/(q-1)}+\delta}, \tag{15}$$

where, $E$ is the energy of quantum, and $\mu_B$ is its baryochemical potential. $\delta = 1$ for fermions, and $\delta = -1$ for Bosons. In the limit of $q \to 1$, it reduces to the conventional Fermi-Dirac or Bose-Einstein distributions. Hence, the value of $q$ reflects the discrepancies of nonextensive statistics from conventional one. Known from Eq. (15), the average energy of quantum in the state with temperature $T$ reads

$$\bar{E}_q = \frac{m_T \cosh(y-y_F)}{\{1+[(q-1)(m_T\cosh(y-y_F)-\mu_B)]/T\}^{1/(q-1)}+\delta}, \tag{16}$$

where, $y$ is the rapidity of quantum, $m_T = \sqrt{p_T^2 + m^2}$ is its transverse mass with rest mass $m$ and transverse momentum $p_T$.

**(2) The rapidity distributions of charged particles in the state of fluid**

In terms of Khalatnikov potential $\chi$, the rapidity distributions of charged particles in the state of fluid can be written as [42]

$$\frac{dN}{dy_F} = \frac{Q_0 c_0}{2} A(b) \left( \cosh y \frac{dz}{dy_F} - \sinh y \frac{dt}{dy_F} \right), \tag{17}$$

where

$$A(b) = 2r^2 \arccos\frac{b}{2r} - b\sqrt{r^2 - \left(\frac{b}{2}\right)^2}$$

is the area of overlap region of collisions, $b$ is impact parameter, and $r$ is the radius of colliding nucleus. From Eqs. (8) and (9), the expression in the round brackets in Eq. (17) becomes

$$\cosh y \frac{dz}{dy_F} - \sinh y \frac{dt}{dy_F}$$
$$= \frac{1}{T}c_s^2 \frac{\partial}{\partial \omega}\left(\chi + \frac{\partial \chi}{\partial \omega}\right)\cosh(y-y_F) - \frac{1}{T}\frac{\partial}{\partial y_F}\left(\chi + \frac{\partial \chi}{\partial \omega}\right)\sinh(y-y_F). \tag{18}$$

**(3) The transverse momentum distributions of charged particles produced in heavy ion collisions**

Along with the expansions of hadronic matter, its temperature becomes even lower. As the temperature drops to kinetic freeze-out temperature $T_f$, the inelastic collisions among hadronic matter stop. The yields of identified hadrons keep unchanged becoming the measured results. According to Cooper-Frye scheme [42], the invariant multiplicity distributions of charged particles take the form as [18, 42, 43]

$$\frac{d^2N}{2\pi p_T\, dy\, dp_T} = \frac{1}{(2\pi)^3} \int_{-y_h(\omega_f)}^{y_h(\omega_f)} \left(\frac{dN}{dy_F} \times \bar{E}_q\right)\bigg|_{T=T_f} dy_F, \tag{19}$$

where $\omega_f = \ln(T_0/T_f)$, and the integrand takes values at the moment of $T = T_f$. Substituting $\chi$ in Eq. (18) by the $\chi_h$ of Eq. (12), it becomes

$$\left(\cosh y \frac{dz}{dy_F} - \sinh y \frac{dt}{dy_F}\right)\bigg|_{T=T_f}$$
$$= \frac{1}{T_f}(\beta_h c_h)^2 S(\omega_f)\left[B(\omega_f, y_F)\sinh(y - y_F) + C(\omega_f, y_F)\cosh(y - y_F)\right], \tag{20}$$

where

$$B(\omega_f, y_F) = \frac{\beta_h y_F}{\lambda(\omega_f, y_F)}\left\{\frac{\beta_h c_h y_h(\omega_f)}{\lambda(\omega_f, y_F)} I_0[\lambda(\omega_f, y_F)] + \left[\frac{\beta_h + 1}{\beta_h} - \frac{2\beta_h c_h y_h(\omega_f)}{\lambda^2(\omega_f, y_F)}\right] I_1[\lambda(\omega_f, y_F)]\right\}, \tag{21}$$

$$C(\omega_f, y_F) = \left\{\frac{\beta_h+1}{\beta_h} + \frac{[\beta_h c_h y_h(\omega_f)]^2}{\lambda^2(\omega_f, y_F)}\right\} I_0[\lambda(\omega_f, y_F)]$$
$$+ \frac{1}{\lambda(\omega_f, y_F)}\left\{\frac{y_h(\omega_f)}{c_h} + 1 - \frac{2[\beta_h c_h y_h(\omega_f)]^2}{\lambda^2(\omega_f, y_F)}\right\} I_1[\lambda(\omega_f, y_F)] \tag{22}$$

where $I_1$ is the 1st order modified Bessel function.

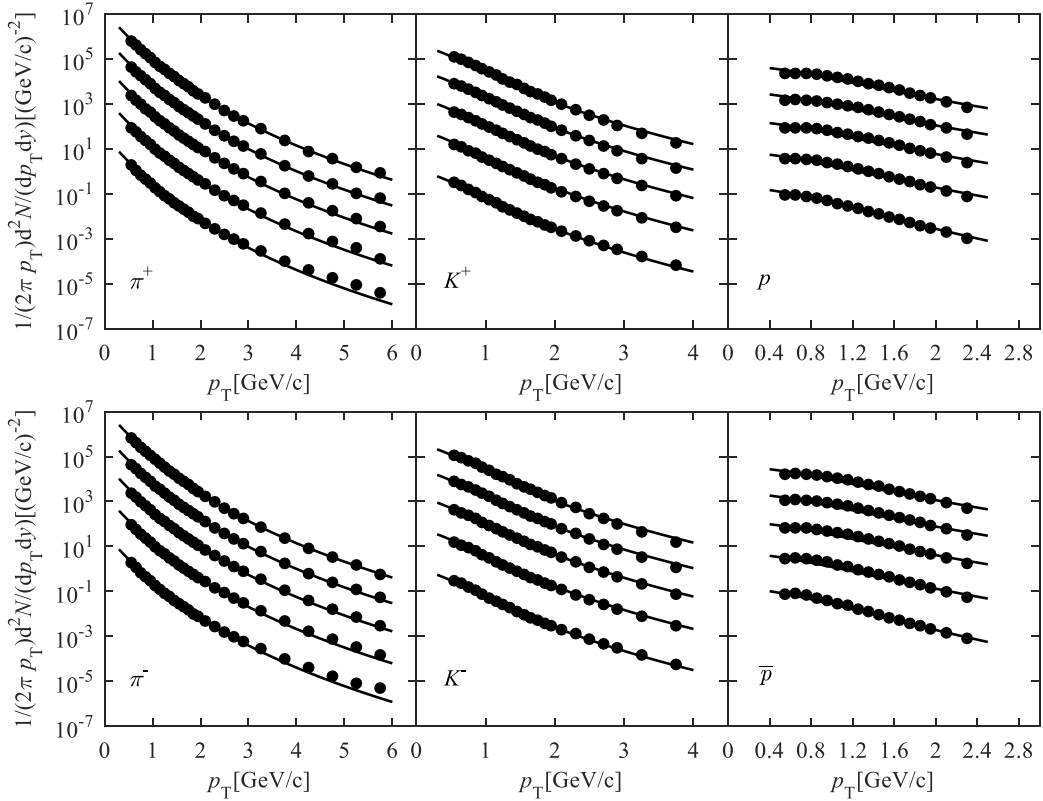

FIGURE 1: The invariant yields of $\pi^\pm$, $K^\pm$, and $p(\bar{p})$ as a function of $p_T$ in Au+Au collisions at $\sqrt{s_{NN}} = 200$ GeV. The solid dots represent the experimental data of the PHENIX Collaboration [3]. The solid curves are the results calculated from Eq. (19). The centrality cuts counted from top to bottom in each panel are 0-10%($\times 10^4$), 10-20%($\times 10^3$), 20-40%($\times 10^2$), 40-60%($\times 10^1$), 60-92%($\times 10^0$), respectively.

By using Eqs. (17) and (19)-(22), we can obtain the transverse momentum distributions of identified charged particles as shown in figures 1 and 2.

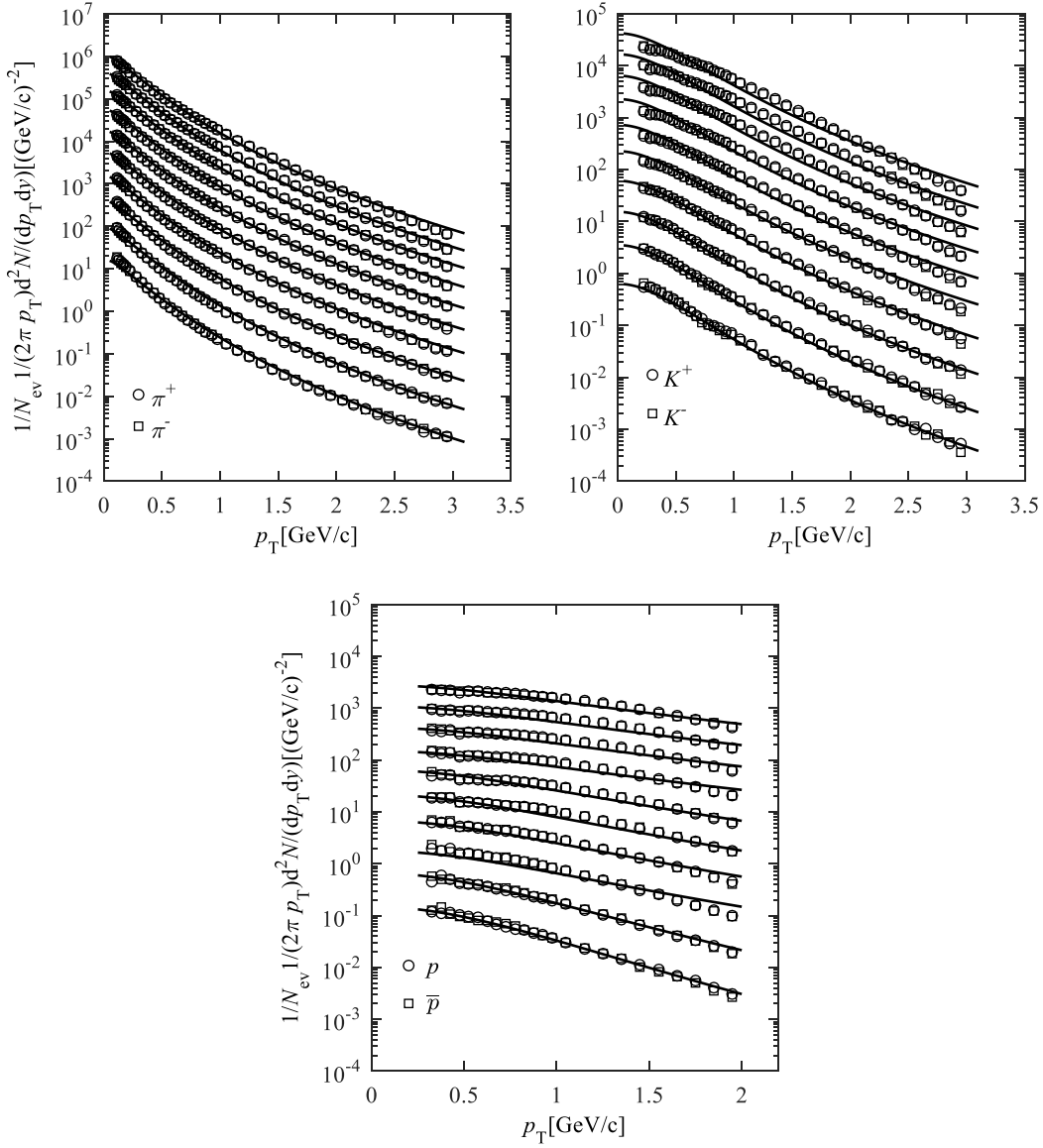

FIGURE 2: The invariant yields of $\pi^{\pm}$, $K^{\pm}$, and $p(\bar{p})$ as a function of $p_T$ in Pb+Pb collisions at $\sqrt{s_{NN}} = 2.76$ TeV. The circles and squares represent the experimental data of the ALICE Collaboration [4]. The solid curves are the results calculated from Eq. (19). The centrality cuts counted from top to bottom in each panel are 0-5%($\times 2^9$), 5-10%($\times 2^8$), 10-20%($\times 2^7$), 20-30%($\times 2^6$), 30-40%($\times 2^5$), 40-50%($\times 2^4$), 50-60%($\times 2^3$), 60-70%($\times 2^2$), 70-80%($\times 2^1$), and 80-90%($\times 2^0$), respectively.

Figure 1 shows the invariant yields of $\pi^{\pm}$, $K^{\pm}$, and $p(\bar{p})$ as a function of $p_T$ in Au+Au collisions at $\sqrt{s_{NN}} = 200$ GeV. Figure 2 shows the same distributions in Pb+Pb collisions at $\sqrt{s_{NN}} = 2.76$ TeV. The solid dots represent the experimental data [3, 4]. The solid curves are the results calculated from Eq. (19). It can be seen that the theoretical results are in good agreement with the experimental data for $\pi^{\pm}$, $K^{\pm}$ in the whole measured $p_T$ region. For $p(\bar{p})$, the theoretical model works well in the region up to about $p_T \leq 2.0$ GeV/c. Beyond this region, the deviation appears as shown in figure 3, which presents the fittings for $p(\bar{p})$ in both the most peripheral collisions for $p_T$ up to about 4 GeV/c.

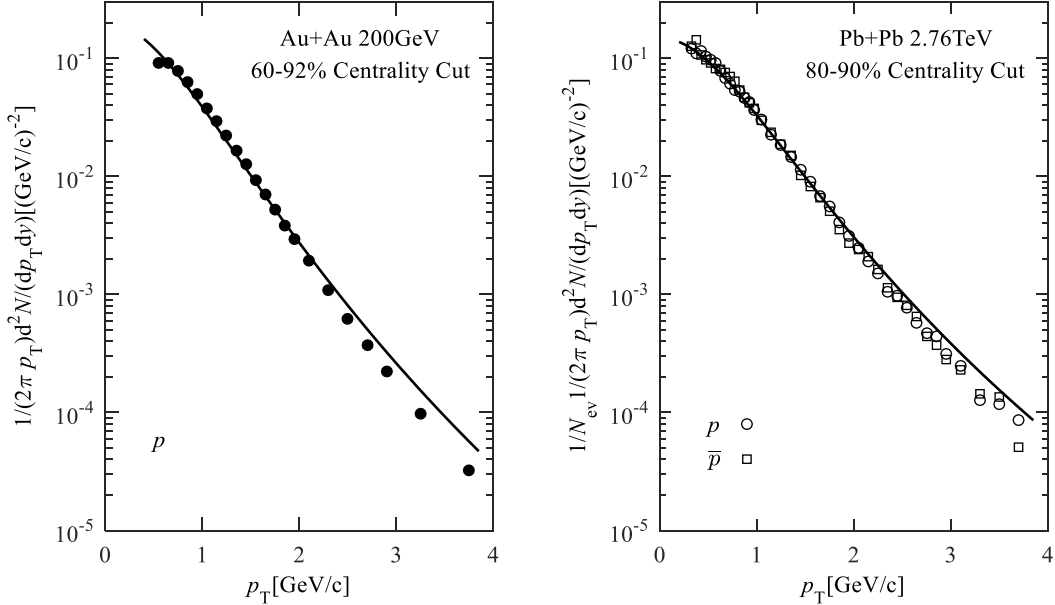

FIGURE 3: The invariant yields of $p(\bar{p})$ as a function of $p_T$ in 60-92% Au+Au collisions at $\sqrt{s_{NN}} = 200$ GeV (left panel) and in 80-90% Pb+Pb collisions at $\sqrt{s_{NN}} = 2.76$ TeV (right panel). The meanings of solid dots, circles, squares, and solid curves are the same as those in figures 1 and 2.

In analyses, the sound speed in hadronic state takes the value of $c_h = 0.35$ [43, 44-46]. The critical temperature takes the value of $T_c = 0.18$ GeV [47]. The kinetic freeze-out temperature $T_f$ takes the values of 0.12 GeV for pions, kaons and 0.13 GeV for protons, respectively, from the investigations of Ref. [8], which also shows that the baryochemical potential $\mu_B$ is about 0.019 GeV in Au+Au collisions. For Pb+Pb collisions, $\mu_B$ takes the value of $\mu_B = 0$ owing to the fact that the yields of particles and antiparticles are equal in such collisions [4]. The initial temperature in central Au+Au collisions takes the same value of $T_0 = 0.7$ GeV as that used in Ref. [43]. For central Pb+Pb collisions, $T_0$ takes the value of $T_0 = 6.5$ GeV referring to that used in Ref. [48]. Tables 1 and 2 list the values of $T_0$, $q$ and $Q_0$ in different centrality cuts. It can be seen that $T_0$ decreases with increasing centrality cuts. $q$ is slightly larger than 1 for different kinds of charged particles. It is almost irrelevant to centrality cuts, while increases with the mass of charged particles on the whole. $Q_0$ is independent of centrality cuts for different kinds of charged particles in Au+Au collisions. The fitted $Q_0$ in table 1 gives

$$\frac{Q_0(\pi^-)}{Q_0(\pi^+)} = 1, \quad \frac{Q_0(K^-)}{Q_0(K^+)} = 0.917, \quad \frac{Q_0(\bar{p})}{Q_0(p)} = 0.7 \text{ (Au+Au 200 GeV)}. \tag{23}$$

These ratios are in good agreement with the relative abundances of particles and antiparticles given in Ref. [3]. This consistency may be due to the fact that the integrand of Eq. (19) is the same for particles and antiparticles in case that $T_f$ takes a common constant for these two kinds of particles. Therefore, $Q_0$ might be proportional to the abundance of corresponding particles. The $Q_0$ in Pb+Pb collisions is also independent of centrality cuts for $\pi^\pm$ and $K^\pm$. While for $p(\bar{p})$, $Q_0$ increases with centrality cuts from semicentral to peripheral collisions. The $Q_0$ listed in table 2 gives the ratios

$$\frac{Q_0(\pi^-)}{Q_0(\pi^+)} = \frac{Q_0(K^-)}{Q_0(K^+)} = \frac{Q_0(\bar{p})}{Q_0(p)} = 1. \tag{24}$$

This is consistent with the above stated fact that, in Pb+Pb collisions at $\sqrt{s_{NN}} = 2.76$ TeV, the yield of charged particles is equal to that of antiparticles.

TABLE 1. The values of $T_0$, $q$, and $Q_0$ in different centrality Au+Au collisions at $\sqrt{s_{NN}} = 200$ GeV.

| Centrality Cuts | $T_0$(GeV) | $q(\pi/K/p)$ | $Q_0(\pi^+/\pi^-)$ | $Q_0(K^+/K^-)$ | $Q_0(p/\bar{p})$ |
|---|---|---|---|---|---|
| 0-10% | 0.70 | 1.08/1.01/1.12 | 0.070/0.070 | 0.024/0.022 | 0.020/0.014 |
| 10-20% | 0.70 | 1.08/1.01/1.12 | 0.070/0.070 | 0.024/0.022 | 0.020/0.014 |
| 20-40% | 0.69 | 1.08/1.01/1.12 | 0.070/0.070 | 0.024/0.022 | 0.020/0.014 |
| 40-60% | 0.68 | 1.08/1.01/1.11 | 0.070/0.070 | 0.024/0.022 | 0.020/0.014 |
| 60-92% | 0.67 | 1.08/1.09/1.09 | 0.070/0.070 | 0.024/0.022 | 0.020/0.014 |

TABLE 2: The values of $T_0$, $q$, and $Q_0$ in deferent centrality Pb+Pb collisions at $\sqrt{s_{NN}} = 2.76$ TeV

| Centrality Cuts | $T_0$(GeV) | $q(\pi/K/p)$ | $Q_0(\times 10^4)$ $(\pi^+/\pi^-)$ | $Q_0(\times 10^4)$ $(K^+/K^-)$ | $Q_0(\times 10^4)$ $(p/\bar{p})$ |
|---|---|---|---|---|---|
| 0-5% | 6.5 | 1.11/1.13/1.26 | 0.750/0.750 | 0.200/0.200 | 0.025/0.025 |
| 5-10% | 6.4 | 1.11/1.13/1.26 | 0.750/0.750 | 0.200/0.200 | 0.025/0.025 |
| 10-20% | 6.2 | 1.11/1.13/1.26 | 0.750/0.750 | 0.200/0.200 | 0.025/0.025 |
| 20-30% | 5.9 | 1.11/1.13/1.26 | 0.750/0.750 | 0.200/0.200 | 0.025/0.025 |
| 30-40% | 5.4 | 1.11/1.13/1.20 | 0.750/0.750 | 0.200/0.200 | 0.060/0.060 |
| 40-50% | 4.9 | 1.11/1.13/1.18 | 0.750/0.750 | 0.200/0.200 | 0.080/0.080 |
| 50-60% | 4.3 | 1.11/1.13/1.18 | 0.750/0.750 | 0.200/0.200 | 0.090/0.090 |
| 60-70% | 3.7 | 1.11/1.12/1.18 | 0.750/0.750 | 0.200/0.200 | 0.090/0.090 |
| 70-80% | 3.1 | 1.11/1.11/1.12 | 0.750/0.750 | 0.200/0.200 | 0.300/0.300 |
| 80-90% | 2.4 | 1.10/1.11/1.10 | 0.750/0.750 | 0.200/0.200 | 0.500/0.500 |

## 4. Conclusions

By introducing nonextensive statistics, we employ the relativistic hydrodynamics including phase transition to discuss the transverse momentum distributions of charged particles produced in Au+Au collisions at $\sqrt{s_{NN}} = 200$ GeV and in Pb+Pb collisions at $\sqrt{s_{NN}} = 2.76$ TeV. The model contains a rich information about the transport coefficients of fluid, such as the initial temperature $T_0$, the critical temperature $T_c$, the kinetic freeze-out temperature $T_f$, the baryochemical potential $\mu_B$, the sound speed in sQGP state $c_0$ and the sound speed in hadronic state $c_h$. Except for $T_0$, the other five parameters take the values either from widely accepted theoretical results or from experimental measurements. As for $T_0$, there are no acknowledged values so far. In this paper, $T_0$ takes the values from other studies for the most central collisions, and for the rest centrality cuts, $T_0$ is determined by tuning the theoretical results to experimental data. The present investigations show the conclusions as follows:

(1) The theoretical model can give a good description to the experimental data in Au+Au collisions at $\sqrt{s_{NN}} = 200$ GeV and in Pb+Pb collisions at $\sqrt{s_{NN}} = 2.76$ TeV for $\pi^{\pm}$, $K^{\pm}$ in the whole measured transverse momentum region, and for $p(\bar{p})$ in the region of $p_T \leq 2.0$ GeV/c.

(2) The fitted $q$ is close to 1. It might mean that the difference between nonextensive statistics and conventional statistics is small. However, it is this small difference that plays an essential role in extending the fitting region of $p_T$.

(3) $Q_0$ is independent of centrality cuts for different charged particles in Au+Au collisions at $\sqrt{s_{NN}} = 200$ GeV. For Pb+Pb collisions, $Q_0$ is irrelevant to centrality cuts for $\pi^{\pm}$ and $K^{\pm}$, while increases from semicentral to peripheral collisions for $p(\bar{p})$.

(4) The methods cannot describe the experimental measurements for $p(\bar{p})$ in the region of $p_T \geq 2.0$ GeV/c for the both kinds of collisions. This might be caused by the hard scattering process [49]. To improve the fitting conditions, the results from perturbative QCD should be taken into account.

## Conflict of Interests

The authors declare that there is no conflict of interests regarding the publication of this paper.

## Acknowledgments

This work is supported by the Shanghai Key Lab of Modern Optical System.